\documentclass{llncs}
\usepackage{subscript} 

\usepackage{makeidx}  
\usepackage{hyperref} 

\usepackage{algorithmic}
\usepackage{caption}
\usepackage{subcaption}

\usepackage{times}
\usepackage{graphicx,verbatim,color,epsfig,amsmath}
\usepackage[ruled,linesnumbered,vlined]{algorithm2e}
\usepackage{ifthen,array}
\usepackage{verbatim,color,latexsym,amssymb,amsmath,multirow,setspace}
\usepackage{xspace}
\usepackage{setspace}
\usepackage{textcomp}
\usepackage{stmaryrd}
\usepackage{hyperref}

\usepackage{enumitem}
\usepackage{caption} 

\usepackage[table]{xcolor}
\definecolor{lightgray}{gray}{0.925}

\usepackage{tikz}

\newenvironment{Proof}{\noindent{\em Proof.~}}{\hfill$\Box$\\[-0.15cm]}

\newcommand{\todoF}[2]{}

\newcommand{\fml}[1]{{\mathcal{#1}}}

\newcommand{\elplus}{$\fml{EL}^{+}$\xspace}
\newcommand{\elplain}{$\fml{EL}$\xspace}
\DeclareMathOperator*{\tbox}{\fml{T}}
\DeclareMathOperator*{\tsbox}{\fml{S}}
\DeclareMathOperator*{\interp}{\fml{I}}
\DeclareMathOperator*{\phiallt}{\phi^{all}_{\mathcal{T}(po)}}
\DeclareMathOperator*{\antec}{\mathsf{ant}}
\DeclareMathOperator*{\gener}{\mathsf{gen}}
\DeclareMathOperator*{\ncdef}{\mathsf{N}_{\mathsf{C}}}
\DeclareMathOperator*{\nrdef}{\mathsf{N}_{\mathsf{R}}}
\DeclareMathOperator*{\pcdef}{\mathsf{PC}_{\mathcal{T}}}
\DeclareMathOperator*{\prdef}{\mathsf{PR}_{\mathcal{T}}}
\DeclareMathOperator*{\eltosat}{\fml{EL}^{+}2\mathsf{SAT}}

\SetKwData{cores}{$\mathcal{C}$}

\definecolor{gray}{rgb}{.4,.4,.4}
\definecolor{midgrey}{rgb}{0.5,0.5,0.5}
\definecolor{darkgrey}{rgb}{0.3,0.3,0.3}
\definecolor{darkred}{rgb}{0.7,0.1,0.1}
\definecolor{darkblue}{rgb}{0.1,0.1,0.5}
\definecolor{defseagreen}{cmyk}{0.69,0,0.50,0}

\newcounter{Comment}[Comment]
\setitemize{noitemsep,topsep=3pt,parsep=2pt,partopsep=2pt}
\setenumerate{noitemsep,topsep=3pt,parsep=2pt,partopsep=2pt}

\usetikzlibrary{shapes,arrows}

\hypersetup{%
    bookmarks=false,    
    pdftitle={},        
    pdfauthor={Joao Marques-Silva},                    
    pdfsubject={TeX and LaTeX},                        
    pdfkeywords={TeX, LaTeX, graphics, images}, 
    colorlinks=true,       
    linkcolor=blue,        
    citecolor=blue,        
    filecolor=black,       
    urlcolor=blue,         
    linktoc=page           
}

\AtBeginDocument{%

}

\begin{document}

\title{Towards Efficient Axiom Pinpointing of $\fml{EL}^+$ Ontologies}
\titlerunning{SAT-Based Debugging of $\fml{EL}^+$ Ontologies}  
%
\author{M.\ Fareed Arif\inst{1} \and Joao Marques-Silva\inst{1,2}}
\authorrunning{Arif \& Marques-Silva} 
\institute{
CASL, University College Dublin, Ireland\\
\email{muhammad.arif.1@ucdconnect.ie,jpms@ucd.ie}
\and
INESC-ID, IST, ULisboa, Portugal\\
}

\maketitle              

\setcounter{footnote}{0}

%
\begin{abstract}
The $\fml{EL}$ family of Description Logics (DLs) has been the subject
of interest in recent years. On the one hand, these DLs are tractable,
but fairly inexpressive. On the other hand, these DLs can be used for
designing different classes of ontologies, most notably ontologies
from the medical domain.
Unfortunately, building ontologies is error-prone. As a result, 
inferable subsumption relations among concepts may be unintended. 
In recent years, the problem of axiom pinpointing has been studied
with the purpose of providing minimal sets of axioms that explain
unintended subsumption relations.
For the concrete case of $\fml{EL}$ and $\fml{EL}^{+}$, the most
efficient approaches consist of encoding the problem into
propositional logic, specifically as a Horn formula, which is then
analyzed with a dedicated algorithm.
This paper builds on this earlier work, but exploits the important
relationship between minimal axioms sets and minimal unsatisfiable
subformulas in the propositional domain.
In turn, this relationship allows applying a vast body of recent work
in the propositional domain to the concrete case of axiom pinpointing
for $\fml{EL}$ and its variants.
From a practical perspective, the algorithms described in this paper
are often several orders of magnitude more efficient that the current
state of the art in axiom pinpointing for the $\fml{EL}$ family of
DLs.
\end{abstract}
%
%

\section{Introduction}

Axiom pinpointing denotes the problem of computing one minimal axiom
set (denoted {\em MinA}), which explains a subsumption relation in an
ontology~\cite{schlobach-ijcai03}.
Axiom pinpointing for different description logics (DLs) has been
studied extensively over the last
decade~\cite{schlobach-ijcai03,parsia-www05,meyer-aaai06,baader-ijcar06,baader-ki07,parsia-iswc07,parsia-jws07,schlobach-jar07,baader-krmed08,sebastiani-cade09,baader-jlc10,meyer-rr11,nguyen-dl12,sebastiani-jair14}, 
but some of the algorithms used can be traced back to the mid
90s~\cite{baader-jar95}.
Description of logics of interest have included $\mathcal{ALC}$,
$\mathcal{SHIF}$, $\mathcal{SHOIN}$, in addition to the \elplain
family of lightweight DLs.
More recent work has focused on the tractable, 
albeit fairly inexpressive, 
$\fml{EL}$ family of description logics (DLs). The reason for this
interest is that the $\fml{EL}$ family of DLs finds important
applications, that include designing of medical ontologies.
Besides computing one MinA, axiom pinpointing is also concerned with
computing all MinAs~\cite{sebastiani-cade09,sebastiani-jair14} or
computing MinAs on demand~\cite{baader-ki07}\footnote{Following
  existing nomenclature, MinA denotes a single axiom set, whereas
  MinAs denotes multiple axiom
  sets~\cite{baader-ki07,sebastiani-cade09}.}.

Original work on axiom pinpointing for the \elplain family of DLs used
the well-known labeling-based classification
algorithm~\cite{baader-dl07,baader-ki07} to find all MinAs for the
$\fml{EL}$ family of DLs. The proposed
approach~\cite{baader-dl07,baader-ki07} generates a worst-case
exponential propositional size formula, which is then used for
computing all MinAs by finding all the minimal models of this
formula.
More recent work~\cite{sebastiani-cade09,sebastiani-jair14} 
proposed an encoding of the problem of computing all MinAs to a
propositional Horn formulae. As an important additional result, it was
shown that the resulting formulas are exponentially more compact than
earlier work in the worst case.
In addition, this work proposed dedicated algorithms for computing
MinAs, based on propositional satisfiability (SAT) solving, but
exploiting techniques used in AllSMT
algorithms~\cite{nieuwenhuis-cav06}. Although effective at computing
MinAs, these dedicated algorithms often fail to enumerate all MinAs to
completion, or proving that no additional MinAs exist.
Nevertheless, the practical application of \elplus in medical
ontologies and the need for axiom pinpointing motivate more efficient
approaches to be developed. 

The main contribution of our work is to show that the computation of
MinAs can be related with the extraction of minimal unsatisfiable
subformulas (MUS) of the Horn formula
encoding proposed in earlier
work~\cite{sebastiani-cade09,sebastiani-jair14}. More concretely, a 
subformula of this encoding is an MUS if and only if it represents one
MinA~\cite{sebastiani-cade09,sebastiani-jair14}. Although this
connection is straightforward in hindsight, we point out that it had
not been investigated since the work was first published in
2009~\cite{sebastiani-cade09}.
The relationship between MUSes and MinAs allows tapping on the large
recent body of work on extracting MUSes, but also on Minimal Correction
Subsets (MCSes), as well as their minimal hitting set
relationship~\cite{reiter-aij87,lozinskii-jetai03,stuckey-padl05,liffiton-jar08,blms-aicomm12,liffiton-cpaior13,pms-aaai13,mshjpb-ijcai13},
which for the propositional case allows exploiting the performance of
modern SAT solvers. The relationship also allows exploring the vast
body of recent work on solving maximum satisfiability
(MaxSAT)~\cite{ansotegui-aij13,mhlpms-cj13} and on enumerating MaxSAT
solutions~\cite{mlms-hvc12}.
The main practical consequences of this insight is that by exploiting
the Horn formulae encoding proposed in earlier
work~\cite{sebastiani-cade09,sebastiani-jair14} we are able to compute
the set of MinAs for the vast majority of existing problem instances,
and most often with many orders of magnitude performance improvements
over what is currently the state of the
art~\cite{sebastiani-cade09,sebastiani-jair14}.

The paper is organized as follows. 
\autoref{sec:prelim} overviews essential definitions and introduces
the notation used throughout the paper.
\autoref{sec:relw} overviews existing work on axiom pinpointing for
the $\fml{EL}$ family of DLs.
\autoref{sec:enumus} briefly overviews recent work on MUS enumeration,
detailing the approach used in the paper.
\autoref{sec:el2mcs} relates SAT-based axiom pinpointing in
$\fml{EL}^{+}$ with MUS extraction and enumeration for propositional
formulae, and summarizes the organization of a new \elplus axiom
pinpointing tool, EL2MCS.
Experimental results on well-known problem instances are analyzed
in~\autoref{sec:res}.
Finally, the paper concludes in~\autoref{sec:conc}.

%

\section{Preliminaries} \label{sec:prelim}

This section introduces the notation and background material used
throughout the paper, both related with description logics, namely the
\elplain family of DLs, and with propositional satisfiability.
\autoref{ssec:elp} briefly overviews
\elplus. Afterwards,~\autoref{ssec:ap-elp} summarizes work on axiom
pinpointing in \elplus. Finally,~\autoref{ssec:sat} reviews
SAT-related definitions.

\subsection{Lightweight Description Logic $\fml{EL}^{+}$} \label{ssec:elp}

This section follows standard descriptions of
\elplus~\cite{baader-ki07,sebastiani-cade09}.
\elplus belongs to the \elplain family of lightweight description
logics.
Starting from a set $\ncdef$ of {\em concept names} and a set $\nrdef$
of {\em role names}, {\em concept descriptions} in \elplus are defined 
inductively using the {\em constructors} shown in the top part
of~\autoref{tab:eldef}.
(As standard in earlier work, uppercase letters $X$, $X_i$, $Y$, $Y_i$
denote generic concepts, uppercase letters $C$, $C_i$, $D$, $D_i$,
$E$, $E_i$ denote concept names and lowercase letters $r$, $r_i$, $s$
denote role names.)
A {\em TBox} (or {\em ontology}) in \elplus is a finite set of {\em
  general concept inclusion} (GCI) and {\em role inclusion} (RI)
axioms. The syntax of GCIs and RIs is shown in the bottom part
of~\autoref{tab:eldef}.
For a TBox $\tbox$, $\pcdef$ denotes the set of {\em primitive
  concepts} of $\tbox$, representing the smallest set of concepts that
contain: (i) the top concept $\top$; and (ii) all the concept names
used in $\tbox$.
$\prdef$ denotes the set of {\em primitive roles} of $\tbox$,
representing the set of all role names used in $\tbox$.
In addition, and for convenience, $X\equiv Y$ corresponds to the two
GCIs $X\sqsubseteq Y$ and $Y\sqsubseteq X$.
Finally, and throughout the paper, $\fml{A}$ denotes a set of
assertions.

The semantics of \elplus is defined in terms of {\em
  interpretations}. An interpretation $\fml{I}$ is a tuple
$(\Delta^{\interp},\cdot^{\interp})$, where $\Delta^{\interp}$
represents the domain, i.e.\ a non-empty set of individuals, and the
interpretation function $\cdot^{\interp}$ maps each concept name
$C\in\ncdef$ to a set $C^{\interp}\in\Delta^{\interp}$, and maps each
role name $r\in\nrdef$ to a binary relation $r^{\interp}$ defined on
$\Delta^{\interp}$,
i.e.\ $r^{\interp}\subseteq\Delta^{\interp}\times\Delta^{\interp}$.
%
The third column of~\autoref{tab:eldef} details the inductive
definitions of $\cdot^{\interp}$ for arbitrary concept descriptions.
An interpretation $\fml{I}$ is a {\em model} of a TBox $\fml{T}$ if
and only if the conditions in semantics (third) column
of~\autoref{tab:eldef} are respected for every GCI and RI axiom in
$\fml{T}$.

The main inference problem for \elplus is the subsumption problem:
\begin{definition}[Concept Subsumption]
  Let $C,D$ represent two \elplus concept descriptions and let $\fml{T}$
  represent an \elplus TBox. $C$ is subsumed by $D$ w.r.t. $\fml{T}$
  (denoted $C\sqsubseteq_{\fml{T}}D$) if $C^{\interp}\subseteq
  D^{\interp}$ in every model $\fml{I}$ of $\fml{T}$.
\end{definition}

\begin{table}[t]
  \rowcolors{1}{lightgray}{}
  \setlength{\tabcolsep}{0.375em}
  \renewcommand{\arraystretch}{1.25}
  \begin{center}
    {\small
      \begin{tabular}{|l|l|l|} \hline
        & Syntax & Semantics \\ \hline\hline
        top & $\top$ & $\Delta^{\interp}$ \\ \hline
        conjunction & $X\sqcap Y$ & $X^{\interp}\cap Y^{\interp}$ \\ \hline
        existential restriction & $\exists r.X$ & 
        $\{ x\in\Delta^{\interp}\,|\,y\in\Delta^{\interp}:(x,y)\in r^{\interp}\land y\in X^{\interp}\}$ \\ \hline \hline
        general concept inclusion & $X\sqsubseteq Y$ & $X^{\interp}\subseteq Y^{\interp}$ \\ \hline
        role inclusion & $r_1\circ\cdots\circ r_n\sqsubseteq s$ & $r_1^{\interp}\circ\cdots\circ r_n^{\interp}\subseteq s^{\interp}$ \\ \hline
      \end{tabular}
    }
    \caption{Syntax and semantics of \elplus} \label{tab:eldef}
  \end{center}
\end{table}

Subsumption algorithms assume that a given TBox $\tbox$ is
normalized~\cite{baader-ijcai05}.
Normalization can be viewed as the process of breaking complex GCIs
into simpler ones. It is well-known that a TBox $\tbox$ can be
normalized in linear time~\cite{baader-ijcai05}.
Moreover, given a normalized TBox, concept subsumption can be
determined in polynomial time~\cite{baader-ijcai05}.

\begin{theorem}[Theorem~1 in~\cite{baader-ki07}]
  Given a TBox $\fml{T}$ in normal form, the subsumption algorithm
  runs in polynomial time on the size of $\fml{T}$.
\end{theorem}

A classification of TBox $\tbox$ represents all subsumption relations
between concept names in $\tbox$. Similarly to subsumption, and
because of the worst-case quadratic number of subsumption relations,
classification can be determined in polynomial time on the size of
$\tbox$~\cite{baader-ijcai05,baader-ki07}.

\begin{example}[GALEN-based Medical Ontology Example] \label{ex:ontol1}
Let us consider  an  $\fml{EL}^+$  medical ontology adapted from the
GALEN medical ontology~\cite{rector-97}, and shown
in~\autoref{table-example-1}.
The ontology expresses a medical condition in which endocarditis is
classified as a heart disease (i.e., Endocarditis $\sqsubseteq$
HeartDisease). This disease  occurs due to a bacteria that damages
endocardium, a tissue, that provides a protection to the heart
valves. 
The table shows both the ontoloy and its classification.
The normalized ontology consists of 11 GCI's and 2 role inclusion
axioms (i.e.,
$\text{contIn}\circ\text{contIn}\sqsubseteq\text{contIn}$ and
$\text{hasLoc}\circ\text{contIn}\sqsubseteq\text{hasLoc}$). Only
assertions are added to the assertion axiom set
$\fml{A}_{\text{MG}}$.
Regarding the classification of this ontology, N denotes the reference
to the assertion axiom
$\text{Endocarditis}\sqsubseteq\exists\text{hasLoc}\text{Heart}$, and
serves to simplify the notation of assertion axioms set.
\begin{table}[t]
\centering
\renewcommand{\arraystretch}{1.05}
\begin{tabular}{|c|}
\hline
$\fml{T}_{\text{MG}}$ :=
$\begin{array} {lcl}\{
\text{Endocarditis}\sqsubseteq\text{Inflammation}\sqcap\exists\text{hasLoc}.\text{Endocardium},\\
\text{Inflammation}\sqsubseteq\text{Disease}\sqcap\exists\text{actsOn}.\text{Tissue}, \\
\text{Endocardium}\sqsubseteq\text{Tissue}\sqcap\exists\text{contIn}.\text{HeartValve},\\
\text{HeartValve}\sqsubseteq\exists\text{contIn}.\text{Heart},\\
\text{HeartDisease}\equiv\text{Disease}\sqcap\exists\text{hasLoc}.\text{Heart},\\
\text{contIn}\circ\text{contIn}\sqsubseteq\text{contIn},\\
\text{hasLoc}\circ\text{contIn}\sqsubseteq\text{hasLoc}\}\\
\end{array}$\\
\hline 
$\fml{A}_{\text{MG}}$ := 
$\begin{array} {lcr}\{
\text{Endocarditis}\sqsubseteq\text{Inflammation}, \\
\text{Inflammation}\sqsubseteq\text{Disease}, \\
\text{Endocardium}\sqsubseteq\text{Tissue}, \\
\text{HeartDisease}\sqsubseteq\text{Disease}, \\
\text{Endocarditis}\sqsubseteq\exists\text{hasLoc}.\text{Endocardium}, \\
\text{Inflammation}\sqsubseteq\exists\text{actsOn}.\text{Tissue}, \\
\text{Endocardium}\sqsubseteq\exists\text{contIn}.\text{HeartValve}, \\
\text{HeartValve}\sqsubseteq\exists\text{contIn}.\text{Heart},\\
\text{HeartDisease}\sqsubseteq\exists\text{hasLoc}.\text{Heart},\\
\text{Disease}\sqcap\text{N}\sqsubseteq\text{HeartDisease},\\
\exists\text{hasLoc}\text{Heart}\sqsubseteq\text{N},\\
\text{Endocarditis}\sqsubseteq\text{Disease},\\ 
\text{Endocarditis}\sqsubseteq\exists\text{actsOn}.\text{Tissue},\\ 
\text{Endocarditis}\sqsubseteq\exists\text{hasLoc}.\text{HeartValve},\\ 
\text{Endocardium}\sqsubseteq\exists\text{contIn}.\text{Heart},\\ 
\text{HeartDisease}\sqsubseteq\text{N},\\ 
\text{Endocarditis}\sqsubseteq\exists\text{hasLoc}\text{Heart}\leftarrow \text{N},\\ 
\text{Endocarditis}\sqsubseteq\text{N},\\  
\textbf{Endocarditis}\sqsubseteq\textbf{HeartDisease}\}\\
\end{array}$\\
\hline
\end{tabular}
\caption{An example $\fml{T}_{\text{MG}}$  
  medical ontology and its classification.}
\label{table-example-1}
\end{table}
\end{example}

\subsection{Axiom Pinpointing in \elplus} \label{ssec:ap-elp}

Axiom pinpointing is the problem of explaining unintended subsumption
relations in  description logics.
As is the case in so many other areas of research, the goal of axiom
pinpointing is to find minimal explanations for unintended subsumption
relations.
The importance of this research topic is illustrated by the large body
of work, targeting different description
logics~\cite{schlobach-ijcai03,parsia-www05,meyer-aaai06,baader-ijcar06,baader-ki07,parsia-iswc07,parsia-jws07,schlobach-jar07,baader-krmed08,sebastiani-cade09,baader-jlc10,meyer-rr11,sebastiani-jair14}.
For the concrete case of the \elplain family of DLs, two main approaches
have been
proposed~\cite{baader-ijcar06,baader-ki07,sebastiani-cade09,sebastiani-jair14}.

Earlier work~\cite{baader-ijcar06,baader-ki07} consists in creating a
pinpointing formula $\phi$, and then enumerating the minimal models of
$\phi$. This is the approach implemented in the CEL
tool~\cite{baader-ijcar06}.
The main drawback of this work is that the pinpointing formula $\phi$
is worst-case exponential on the size of $\tbox$, and enumeration of
minimal models is NP-hard.
In contrast, more recent work showed that concept subsumption can be
encoded to a Horn formula, and that the axiom pinpointing problem can
be solved on this Horn
formula~\cite{sebastiani-cade09,sebastiani-jair14}. This approach is
detailed in~\autoref{sec:relw}.

Throughout this paper, the following standard definition of MinA (and
of nMinA) is assumed.

\begin{definition}[nMinA/MinA]
Let $\tbox$ be an \elplus TBox, and let $C, D\in\pcdef$ be primitive
concept names, with $C\sqsubseteq_{\tbox}D$. 
Let $\tsbox$ be a subset of $\tbox$ be such that
$C\sqsubseteq_{\tsbox}D$.
If $\tsbox$ is such that $C\sqsubseteq_{\tsbox}D$ and
$C\not\sqsubseteq_{\tsbox'}D$ for $\tsbox'\subset\tsbox$, then
$\tsbox$ is a {\em minimal axiom set} (MinA)
w.r.t.\ $C\sqsubseteq_{\tbox}D$.
Otherwise, $\tsbox$ is a non-minimal axiom set (nMinA)
w.r.t.\ $C\sqsubseteq_{\tbox}D$.
\end{definition}

\subsection{Propositional Satisfiability} \label{ssec:sat}

Standard propositional satisfiability (SAT) definitions are
assumed~\cite{sat-handbook09}. This includes standard definitions for
variables, literals, clauses and CNF formulas.
Formulas are represented by $\fml{F}$, $\fml{M}$, $\fml{M}'$, $\fml{C}$
and $\fml{C}'$, but also by $\phi$ and $\psi$.
Horn formulae are such that every clause contains at most one positive
literal. It is well-known that SAT on Horn formulae can be decided in
linear time~\cite{gallier-jlp84}.
The paper explores both MUSes and MCSes of propositional formulae.

\begin{definition}[MUS] \label{def:mus}
$\fml{M}\subseteq\fml{F}$ is a {\em Minimal Unsatisfiable
  Subformula} (MUS) of $\fml{F}$ iff $\fml{M}$ is unsatisfiable and
$\forall_{\fml{M}'\subsetneq\fml{M}}\,\fml{M}'$ is satisfiable.
\end{definition}

\begin{definition}[MUS] \label{def:mcs}
$\fml{C}\subseteq\fml{F}$ is a {\em Minimal Correction
  Subformula} (MCS) of $\fml{F}$ iff $\fml{F}\setminus\fml{C}$ is
  satisfiable and
  $\forall_{\fml{C}'\subsetneq\fml{C}}\,\fml{F}\setminus\fml{C}'$ is
  unsatisfiable.
\end{definition}

A well-known result, which will be used in the paper is the minimal
hitting set relationship between MUSes and MCSes of an unsatisfiable
formula
$\fml{F}$~\cite{reiter-aij87,lozinskii-jetai03,stuckey-padl05,liffiton-jar08}.

\begin{theorem} \label{th:dual}
  Let $\fml{F}$ be unsatisfiable. Then,
  \begin{enumerate}
  \item Each MCS of $\fml{F}$ is a minimal hitting set of the MUSes of
    $\fml{F}$. 
  \item Each MUS of $\fml{F}$ is a minimal hitting set of the MCSes of
    $\fml{F}$. 
  \end{enumerate}
\end{theorem}
As highlighted in~\autoref{sec:enumus}, \autoref{th:dual} forms the
basis of {\em all} existing approaches for enumerating
MUSes~\cite{stuckey-padl05,liffiton-jar08,liffiton-cpaior13,pms-aaai13}.
Moreover, the importance of set duality relationships in combinatorial
optimization is highlighted by recent work~\cite{slaney-ecai14}.


%
%

\section{SAT-Based Axiom Pinpointing in $\fml{EL}^{+}$} \label{sec:relw}

This section reviews EL2SAT, a recently proposed approach for axiom
pinpointing of \elplus ontologies using a Horn formula 
encoding~\cite{sebastiani-cade09,sebastiani-jair14}.
The EL2SAT approach can be divided into two main components. First,
the axiom pinpointing problem is encoded as a Horn formula which is
polynomial on the size of $\tbox$.
Second, the MinAs are computed using a dedicated algorithm, exploiting
ideas from early work on AllSMT~\cite{nieuwenhuis-cav06}.

The EL2SAT axiom pinpointing approach for \elplus can be summarized as
follows~\cite{sebastiani-cade09,sebastiani-jair14}:
\begin{enumerate}[label=\Alph*.]
\item Encode the classification of TBox $\tbox$ by running the
  standard classification algorithm and adding Horn clauses for
  representing every non-trivial axiom or assertion as a set of Horn
  clauses. The resulting Horn formula is denoted $\phi_{\tbox}$.
\item Encode the complete classification DAG of the input normalized
  ontology $\tbox$ as a Horn formula $\phi^{all}_{\tbox}$.
\item Associate a Boolean variable $s_{[a_i]}$ with each assertion
  $a_i$ in the classification of $\tbox$ and create a clause
  $s_{[a_i]}\to\eltosat(a_i)$, where $\eltosat(a_i)$ is the set of
  clauses associated with $a_i$ in $\phi^{all}_{\tbox}$. The resulting
  formula is denoted $\phi_{{\tbox}(so)}$.
\item Encode the {\em pinpointing-only} problem as another Horn
  formula $\phiallt$. For the purposes of this paper, this is the
  formula that is of interest and so the one analyzed in greater
  detail.
\end{enumerate}
The construction of the target Horn formula $\phiallt$ mimics the
construction of the other formulas, in that the classification
procedure is executed. The formula is constructed as follows:
\begin{enumerate}
\item For every RI axiom, create an axiom selector variable
  $s_{[a_i]}$. For trivial GCI of the form $C\sqsubseteq C$ or
  $C\sqsubseteq\top$, $s_{[a_i]}$ is constant \texttt{true}. For each
  non-trivial GCIs, add an axiom selector variable $s_{[a_i]}$.
\item During the execution of the classification algorithm, for every
  application of a rule (concretely $r$) generating some assertion
  $\gener(r)$ (concretely $a_i$), add to $\phiallt$ a clause of the
  form,
  \begin{equation}
    \left(\bigwedge_{a_j\in\antec(r)}s_{[a_j]}\right)\to s_{[a_i]}
  \end{equation}
  where $s_{[a_i]}$ is the selector variable for $a_i$ and $\antec(r)$
  are the antecedents of $a_i$ with respect to rule $r$.
\end{enumerate}
As explained in earlier work, the encoding procedure ensures that each
rule application is applied only once.
Finally, for the concrete case of axiom pinpointing, specify the
assumption list $\{ \neg s_{[C_i\sqsubseteq D_i]} \} \cup \{
s_{[a_i]}\,|\, a_i \in \tbox \}$. 
This assumption list is manipulated
by the dedicated algorithm proposed in earlier
work~\cite{sebastiani-cade09,sebastiani-jair14}.

The following theorem is fundamental for earlier
work~\cite{sebastiani-cade09,sebastiani-jair14}, and is extended in
the next section to related MinAs with MUSes of propositional
formulae.

\begin{theorem}[Theorem~3 in~\cite{sebastiani-jair14}] \label{th:el2sat}
    Given an \elplus TBox $\tbox$, for every $\tsbox\subseteq\tbox$
    and for every pair of concept names $C,D\in\pcdef$,
    $C\sqsubseteq_{\tsbox}D$ if and only if the Horn propositional
    formula,
    \begin{equation}
      \phiallt\land_{ax_i\in\tsbox}(s_{[ax_i]})\land(\neg
      s_{[C\sqsubseteq D]})
    \end{equation}
    is unsatisfiable.
\end{theorem}

\begin{example}[GALEN Medical Ontology SAT Encoding] \label{ex:ppdl1}
For the ontology in~\autoref{ex:ontol1},~\autoref{table-example-b}
illustrates the construction of the formulas described earlier in this
section.
Regarding the $s_{[a_i]}$ variables, the original set is
$\{s_1,\ldots,s_{13}\}$. These represent the axioms in the normalized
ontology. The remaining ones result from the classification
procedure. 
It it important to note that we do not expand $\eltosat(a_i)$ in
$\phi_{\fml{T}_{\text{MG}(so)}}$. Moreover, a selection variable is
assigned to each role inclusion axiom.
\begin{table}
\centering
\renewcommand{\arraystretch}{1.25}
\begin{tabular}{|c|}
\hline 
$\phi_{\fml{T}_{\text{MG}(so)}}$ :=  $\begin{array} {lcl}
s_1\rightarrow \text{Endocarditis}\sqsubseteq\text{Inflammation}\\
s_2\rightarrow \text{Inflammation}\sqsubseteq\text{Disease} \\
s_3\rightarrow \text{Endocardium}\sqsubseteq\text{Tissue} \\
s_4\rightarrow \text{HeartDisease}\sqsubseteq\text{Disease} \\
s_5\rightarrow \text{Endocarditis}\sqsubseteq\exists\text{hasLoc}.\text{Endocardium} \\
s_6\rightarrow \text{Inflammation}\sqsubseteq\exists\text{actsOn}.\text{Tissue} \\
s_7\rightarrow \text{Endocardium}\sqsubseteq\exists\text{contIn}.\text{HeartValve} \\
s_8\rightarrow \text{HeartValve}\sqsubseteq\exists\text{contIn}.\text{Heart}\\
s_9\rightarrow \text{HeartDisease}\sqsubseteq\exists\text{hasLoc}.\text{Heart}\\
s_{10}\rightarrow \text{Disease}\sqcap\text{N}\sqsubseteq\text{HeartDisease}\\
s_{11}\rightarrow \exists\text{hasLoc}\text{Heart}\sqsubseteq\text{N}\\
s_{12}\rightarrow \text{contIn}\circ\text{contIn}\sqsubseteq\text{contIn}\\ 
s_{13}\rightarrow \text{hasLoc}\circ\text{contIn}\sqsubseteq\text{hasLoc}\\ 
s_{14}\rightarrow \text{Endocarditis}\sqsubseteq\text{Disease}\\ 
s_{15}\rightarrow \text{Endocarditis}\sqsubseteq\exists\text{actsOn}.\text{Tissue}\\ 
s_{16}\rightarrow \text{Endocarditis}\sqsubseteq\exists\text{hasLoc}.\text{HeartValve}\\ 
s_{17}\rightarrow \text{Endocardium}\sqsubseteq\exists\text{contIn}.\text{Heart}\\ 
s_{18}\rightarrow \text{HeartDisease}\sqsubseteq\text{N}\\ 
s_{19}\rightarrow\text{Endocarditis}\sqsubseteq\exists\text{hasLoc}\text{Heart}\leftarrow \text{N}\\ 
s_{20}\rightarrow\text{Endocarditis}\sqsubseteq\text{N}\\  
s_{21}\rightarrow \textbf{Endocarditis}\sqsubseteq\textbf{HeartDisease}\\
\end{array}$\\
\hline 
$\phi^{all}_{\fml{T}_{\text{MG}}(po)}$ :=  $\begin{array} {lcl}
\{s_1\land s_2\rightarrow s_4,  \\
s_6\land s_1\rightarrow s_{15},\\
s_{13}\land s_7\land s_5 \rightarrow s_{16},\\
s_{12}\land s_8\land s_7\rightarrow s_{17},\\
s_{11}\land s_9\rightarrow s_{18},\\
s_{13}\land s_{16}\land s_{8}\rightarrow s_{19},\\
s_{13}\land s_{17}\land s_{5}\rightarrow s_{19},\\
s_{11}\land s_{19}\rightarrow s_{20},\\
s_{10}\land s_{14}\land s_{20}\rightarrow s_{21},\\
s_{4}\land s_{21}\rightarrow s_{14},\\
s_{9}\land s_{21}\rightarrow s_{19}\}\\
\end{array}$\\
\hline
$\phi^{all}_{\fml{T}_{\text{MG}}}:=\phi^{all}_{\fml{T}_{\text{MG}}(po)}\land\bigwedge_{1\leq i\leq 13}(s_i)\land(\lnot s_{21})$\\[3pt]
\hline
\text{MinA} := $\{s_1,s_5,s_8, s_{10}, s_{11}, s_{13}\}$\\
\hline
\end{tabular}
\caption{The $\fml{T}_{\text{MG}}$ Horn encoding~\cite{sebastiani-cade09,sebastiani-jair14}.}
\label{table-example-b}
\end{table}
\end{example}

%
%

\section{Enumeration of MUSes} \label{sec:enumus}

The problem of enumerating all the MUSes of an unsatisfiable formula
has been studied in different
settings~\cite{reiter-aij87,lozinskii-jetai03,stuckey-padl05,liffiton-jar08,liffiton-cpaior13,pms-aaai13},
starting with the seminal work of Reiter~\cite{reiter-aij87}.

Although the enumeration of MCSes can be achieved in a number of
different ways~\cite{mshjpb-ijcai13}, the enumeration of MUSes is
believed to be far more challenging. Intuitively, the main difficulty
is how to block one MUS and then use a SAT solver (or some other
decision procedure) to compute the next MUS.
This difficulty with MUS enumeration motivated a large body of work
exploiting a fundamental relationship between MUSes and MCSes. Indeed,
it is well-known (e.g.\ see~\autoref{th:dual}) that MCSes are
minimal hitting sets of MUSes, and MUSes are minimal hitting sets of
MCSes~\cite{reiter-aij87,lozinskii-jetai03,stuckey-padl05,liffiton-jar08}.
As a result, early approaches~\cite{stuckey-padl05,liffiton-jar08} for
enumeration of MUSes were organized as follows: (i) enumerate all
MCSes of a formula; (ii) compute the minimal hitting sets of the MCSes.

One problem with early approaches is that all MCSes need to be
enumerated {\em before} the first MUS is computed. As a result, more
recent work proposed alternative approaches which enable both MUSes
and MCSes to be computed while enumeration of MUSes (and MCSes) takes
place~\cite{liffiton-cpaior13,pms-aaai13}.
Nevertheless, if the number of MCSes is manageable, earlier work is
expected to be more efficient~\cite{liffiton-jar08}.
It should noted that, all existing approaches for MUS enumeration
relate with~\autoref{th:dual}, in that enumeration of MUSes is
achieved by explicitly or implicitly computing the minimal hitting
sets of all MCSes.

Regarding MCS enumeration, two main approaches have been studied. One
exploits MaxSAT enumeration~\cite{liffiton-jar08,mlms-hvc12}. More
recent work proposes dedicated MCS extraction algorithms, also capable
of enumerating MCSes~\cite{mshjpb-ijcai13}.
The approach proposed in this paper exploits MaxSAT enumeration.

%
%

\section{Axiom Pinpointing with MUS Extraction} \label{sec:el2mcs}

This section shows that the computation of MinAs can be related with
MUS enumeration of the Horn formula $\phi^{all}_{\tbox(po)}$.
It then briefly overviews existing approaches for MUS enumeration, and
concludes by summarizing the MUS enumeration approach used in this
paper.

\subsection{MinAs as MUSes} \label{ssec:mina2mus}

Although not explicitly stated, the relation between axiom pinpointing
and MUS extraction has been apparent in earlier
work~\cite{baader-ki07,sebastiani-cade09,sebastiani-jair14}.
Indeed, from the encoding of axiom pinpointing to Horn formula, it is
immediate the following result, that relates~\autoref{th:el2sat}
(Theorem~3 in~\cite{sebastiani-jair14}) with the dedicated algorithm
proposed in earlier work~\cite{sebastiani-cade09,sebastiani-jair14}:

\begin{theorem} \label{th:el2mus}
    Given an \elplus TBox $\tbox$, for every $\tsbox\subseteq\tbox$
    and for every pair of concept names $C,D\in\pcdef$,
    $\tsbox$ is a MinA of $C\sqsubseteq_{\tsbox}D$ if and only if the
    Horn propositional formula,
    \begin{equation} \label{eq:phipo}
      \phiallt\land_{ax_i\in\tsbox}(s_{[ax_i]})\land(\neg s_{[C\sqsubseteq D]})
    \end{equation}
    is minimally unsatisfiable.
\end{theorem}

\begin{Proof}[Sketch]
  By~\autoref{th:el2sat}, $C\sqsubseteq_{\tsbox}D$ if and only if the
  associated Horn formula~\eqref{eq:phipo} is unsatisfiable.
  For a MinA $\tsbox\subseteq\tbox$, minimal unsatisfiability of
  \eqref{eq:phipo} (with $\tbox$ replaced by $\tsbox$) results from
  the MinA computation algorithm proposed in earlier
  work~\cite{sebastiani-cade09,sebastiani-jair14}.
\end{Proof}

Based on~\autoref{th:el2mus} and the MUS enumeration approaches
summarized in~\autoref{sec:enumus}, we can now outline our approach 
based on MUS enumeration.

Earlier work~\cite{sebastiani-cade09,sebastiani-jair14} explicitly
enumerates assignments to the $s_{[ax_i]}$ variables in a
AllSMT-inspired approach~\cite{nieuwenhuis-cav06}.
In contrast, our approach is to model the problem has partial maximum 
satisfiability (MaxSAT), and enumerate the MUSes of the MaxSAT problem
formulation.

All clauses in $\phiallt$ are declared as hard clauses, i.e.\ they
must be satisfied. Observe that, by construction, $\phiallt$ is
satisfiable.
In addition, the constraint $C\sqsubseteq_{\tbox}D$ is encoded
with another hard clause, namely $(\neg s_{[C\sqsubseteq_{\tbox}D]})$.
Finally, the variable associated with each axiom $ax_i$, $s_{[ax_i]}$
denotes a {\em unit soft clause}. The intuitive justification is that
the goal is to include as many axioms as possible, leaving out a
minimal set which if included will cause the complete formula to be
unsatisfiable.
Thus, each of these sets represents an MCS of the MaxSAT problem
formulation, but also a minimal set of axioms that needs to be dropped
for the subsumption relation not to hold.
MCS enumeration can easily be implemented with a MaxSAT
solver~\cite{liffiton-jar08,mlms-hvc12} or with a dedicated
algorithm~\cite{mshjpb-ijcai13}.
Moreover, we can now use minimal hitting set
dualization~\cite{reiter-aij87,lozinskii-jetai03,stuckey-padl05,liffiton-jar08}
to obtain the MUSes we are looking for, starting from the previously
computed MCSes.
This is the approach implemented in this paper.

\begin{example} \label{ex:mxsat1}
  For the ontology in~\autoref{ex:ontol1}
  and~\autoref{ex:ppdl1},~\autoref{table-example-c} summarizes the
  MaxSAT formulation, as well as an example MUS.
  \begin{table}[t]
    \begin{center}
      \renewcommand{\arraystretch}{1.25}
      \begin{tabular}{|c|} \hline
        $\phi_{\fml{H}} := \{\phi^{all}_{\fml{T}_{\text{MG}}(po)}\}\cup\{\neg s_{21}\}$\\
        $\phi_{\fml{S}} := \{s_1, s_2,\dots,s_{13}\}$\\
        $\Omega := \langle\phi_{\fml{H}}, \phi_{\fml{S}}\rangle$\\ \hline
        \text{MUS} := $\{s_1,s_5,s_8, s_{10}, s_{11}, s_{13}\}$\\ \hline
      \end{tabular}
      \caption{The MaxSAT formulation.} \label{table-example-c}
    \end{center}
  \end{table}
  The hard clauses are given by $\phiallt$ and by $(\neg
  s_{[C\sqsubseteq D]}$. Each positive unit soft clauses is given by
  the selection variable associated with each of the original axioms.
  The EL2MCS approach starts by computing the MCSes, using a
  MaxSAT-based approach, and then computes the MUSes by minimal
  hitting set dualization (e.g.\ see~\autoref{th:dual}).
\end{example}                         

\paragraph{Related Work.}
Reiter's work on computing minimal hitting sets~\cite{reiter-aij87} is
used to find all the justifications in earlier
work~\cite{parsia-iswc07,sb-phd09}, which exploits hitting set trees
(HST). This approach starts by computing a single justification by
using either a standard blackbox method, either a sliding window
deletion based-deletion algorithm (SINGLE\_JUST\_ALG
in~\cite{parsia-iswc07}) or a binary search based  algorithm
(log-extract-mina in~\cite{baader-krmed08}). After finding the first
justification using any of these methods, the algorithm removes axioms
one by one from this justification set and constructs a Hitting Set
Tree (HST) that in turn serves to find all the justifications (using
either SINGLE\_JUST\_ALG or log-extract-mina on each branch of the
HST). The main drawback of this approach is that it does not scale for
large size ontologies. Therefore, this method is only used in conjunction with
reachability-based modules~\cite{baader-krmed08} and when a TBox contains
only a few relevant axioms subject to an entailment. 

\subsection{The EL2MCS Axiom Pinpointing Approach}

\begin{figure}[t]
  \begin{center}
    \scalebox{0.775}{
      \tikzstyle{decision} = [diamond, draw, fill=blue!10,
        text width=4.5em, text badly centered, node distance=3cm, inner sep=0pt]
      \tikzstyle{block} = [rectangle, draw, fill=blue!10,
        text width=7.5em, text centered, rounded corners, minimum height=4.5em]
      \tikzstyle{line} = [draw, thick, -latex']
      \tikzstyle{cloud} = [draw, ellipse,fill=red!10, node distance=3cm,
        minimum height=2em]
      \begin{tikzpicture}[node distance = 4.25cm, auto]
        \node [block] (horn) {Generate Horn formula (EL2SAT)~\cite{sebastiani-cade09,sebastiani-jair14}};
        \node [block, right of=horn] (pmx) {Partial MaxSAT encoding \\(\autoref{ssec:mina2mus})};
        \node [block, right of=pmx] (mcses) {Compute all MCSes (CAMUS2)~\cite{mshjpb-ijcai13}};
        \node [block, right of=mcses] (muses) {Compute all MUSes (CAMUS)~\cite{liffiton-jar08}};
        \path [line] (horn) -- (pmx);
        \path [line] (pmx) -- (mcses);
        \path [line] (mcses) -- (muses);
      \end{tikzpicture}
    }
    \caption{The EL2MCS tool} \label{fig:el2mcs}
  \end{center}
\end{figure}
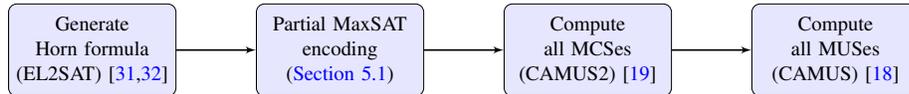

This section summarizes the organization of the EL2MCS tool, that
exploits MUS enumeration for axiom pinpointing.
The organization of EL2MCS is shown in~\autoref{fig:el2mcs}.
The first step is similar to
EL2SAT~\cite{sebastiani-cade09,sebastiani-jair14} in that a
propositional Horn formula is generated.
The next step, however, exploits the ideas in the previous section,
and generates a partial MaxSAT encoding.
As outlined earlier, we can now enumerate the MCSes of the partial
MaxSAT formula. This is achieved with the CAMUS2
tool~\cite{mshjpb-ijcai13}\footnote{Available
  from~\url{http://logos.ucd.ie/web/doku.php?id=mcsls}.}.
The final step is to exploit minimal hitting set dualization for
computing all the MUSes given the set of
MCSes~\cite{liffiton-jar08}. This is achieved with the 
CAMUS tool\footnote{Available
  from~\url{http://sun.iwu.edu/~mliffito/camus/}.}.
It should be observed that, although MCS enumeration uses CAMUS2 (a
modern implementation of the MCS enumerator in
CAMUS~\cite{liffiton-jar08}, capable of handling partial MaxSAT
formulae), alternative MCS enumeration approaches were 
considered~\cite{mshjpb-ijcai13}, not being as efficient.

%
%
\section{Experimental Results} \label{sec:res}
The experiments were performed on an HPC cluster, with dual quad-core
Intel Xeon 3GHz processors, with 32GB of physical memory.
The tools EL2SAT  and EL2MCS were given a timeout of 3600 seconds and
a memory limit of 16GB. EL2MCS uses the partial MaxSAT  instances
generated with the help of EL2SAT Horn propositional encoding
tool\footnote{The $\fml{EL}^+$ to SAT encoder is denoted el2sat\_all,
  whereas the axiom pinpointing tool is el2sat\_all\_mins. These tools
  are available from \url{http://disi.unitn.it/\~rseba/elsat/}.
  Moreover, this site also contains the subsumption query instances
  used in the experiments.}.
The  medical ontologies used in the experimentation are GALEN~\cite{rector-97}, Gene~\cite{ashburner-ng00},
NCI~\cite{sioutos-jbi07} and SNOMED-CT~\cite{spackman-amia97}\footnote{GENE, GALEN and NCI
  ontologies are freely available at
  \url{http://lat.inf.tu-dresden.de/~meng/toyont.html}. The SNOMED-CT
  ontology was requested from IHTSDO under a nondisclosure license
  agreement.}.
We have used the 450 subsumption query instances which are studied in
earlier work~\cite{sebastiani-jair14} (and which are available from
the EL2SAT website).
The experimental results compare exclusively EL2SAT and EL2MCS, since
EL2SAT has recently been shown to consistently outperform
CEL~\cite{baader-ijcar06}, a $\fml{EL}^+$ axiom pinpointing 
tool~\cite{sebastiani-jair14}.
Moreover, unless otherwise stated, the experiments consider the
optimizations proposed in earlier
work~\cite{sebastiani-cade09,sebastiani-jair14}, and exploited in the
EL2SAT tool.
One example is the cone-of-influence (COI) simplification
technique~\cite{sebastiani-jair14}, which enables significant
reductions in the obtained propositional satisfiability instances.
It should be noted that cone-of-influence reduction technique can be
related with $\fml{EL}^+$ reachability-based
modularization~\cite{baader-krmed08}.
\begin{figure}[t]
  \begin{center}
    \hspace*{-0.35cm}
    \includegraphics[scale=0.625]{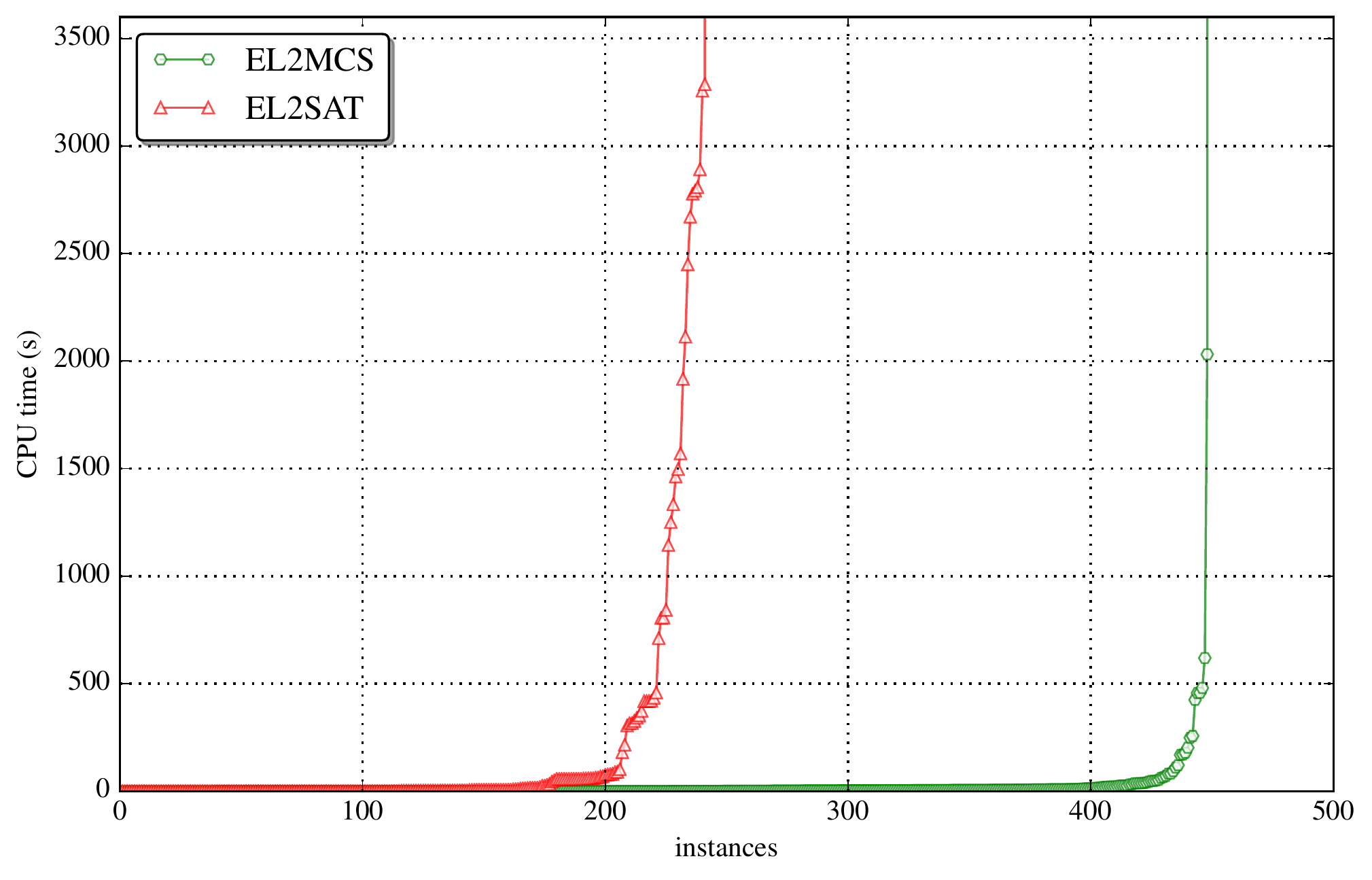}
    \caption{Cactus plot comparing EL2MCS and EL2SAT on all problem
      instances, with COI reduction}
    \label{fig:cactus}
  \end{center}
\end{figure}

\begin{figure}[t]
  \begin{center}
    \includegraphics[scale=0.675]{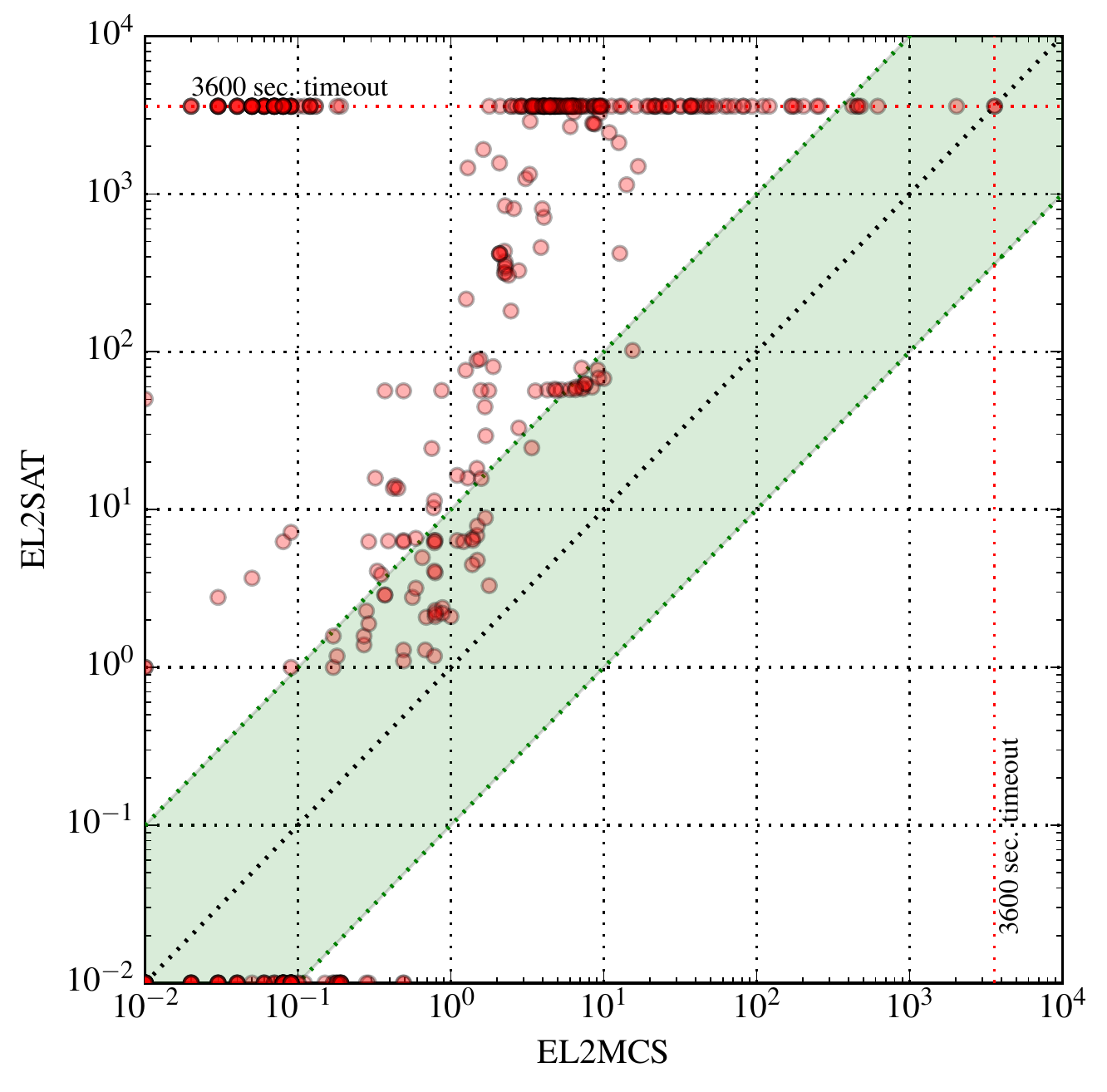}
    \caption{Scatter plot comparing EL2MCS and EL2SAT on all problem
      instances, with COI reduction}
    \label{fig:scatter}
  \end{center}
\end{figure}

\autoref{fig:cactus} shows a cactus plot comparing EL2SAT with EL2MCS.
The performance gap between EL2SAT and EL2MCS is clear and conclusive.
As summarized in~\autoref{tab:stats}, EL2SAT solves 241 out of 450
instances, whereas EL2MCS solves 448 out of 450 instances.
\autoref{fig:scatter} shows a scatter plot comparing the two tools. 
As before, the performance gap between EL2SAT and EL2MCS is clear,
with performance gains that often exceed one order of
magnitude, and that can even exceed three orders of magnitude.
More significantly, for 207 instances (i.e. 209$-$2 out of 450), and
in contrast with EL2MCS, EL2SAT does not terminate within the given
timeout.

\autoref{tab:stats} summarizes the statistics of running the two tools
with and without COI reduction. As can be concluded, COI reduction is
far more relevant for EL2SAT than for EL2MCS. When COI reduction is
used, EL2SAT outperforms EL2MCS on 16.9\% of the instances. It should
be noted that {\em all} of these instances terminate in less that 0.5
seconds for both tools, as can be concluded from
\autoref{fig:scatter}.
In contrast, EL2MCS outperforms EL2SAT on 74.9\% of the instances, and
for 207 (i.e.\ $209-2$) of these, EL2SAT does not terminate.
The statistics further support the conclusion that the extraction of
MUSes provides a far more robust solution than a dedicated algorithm
based on enumeration of subsets and exploiting AllSMT techniques.

\autoref{tab:wins} summarizes the number of times each tool computes
more MUSes than the other tools, independently of being able to prove
the non-existence of additional MUSes within the given timeout.
As can be observed, EL2SAT is actually quite capable of finding most
MUSes, especially when COI reduction is applied. The reason is how the
algorithm is implemented and the preference to promot conflicts.
However, as shown in~\autoref{fig:cactus} and~\autoref{fig:scatter},
in many cases EL2SAT takes significantly more time to compute all
MUSes, and it is often unable to prove the non-existence of additional
MUSes.
These results also indicate that MUSes in the axiom pinpointing
instances considered are usually small. As a result, EL2SAT, which
uses assumptions for the implicit set enumeration of target sets, is
able to compute most MUSes in many cases. In contrast, in situations
where the size of MUSes is larger, EL2SAT would be expected to be
unable to reach a stage where the sets representing MUSes would be
enumerated.

\begin{table*}[t]
  \begin{center}
    \begin{minipage}{0.425\textwidth}
      \begin{center}
        \begin{tabular}{|c|c|c|} \hline
          \rowcolor{lightgray}
          & EL2SAT & EL2MCS \\ \hline\hline
          \# Solved & 241 & 448 \\ \hline
          \% Solved & 53.6\% & 99.6\% \\ \hline
          \# TO     & 209    & 2      \\ \hline
          \% TO     & 46.4\% & 0.4\%  \\ \hline
          \% Wins   & 16.9\% & 74.9\% \\ \hline
        \end{tabular}

        \smallskip {\small (a) With COI reduction}
      \end{center}
    \end{minipage}
    \begin{minipage}{0.125\textwidth}
      \quad
    \end{minipage}
    \begin{minipage}{0.425\textwidth}
      \begin{center}
        \begin{tabular}{|c|c|c|} \hline
          \rowcolor{lightgray}
          & EL2SAT & EL2MCS \\ \hline\hline
          \# Solved & 8 & 447 \\ \hline
          \% Solved & 1.8\% & 99.3\% \\ \hline
          \# TO     & 442    & 3      \\ \hline
          \% TO     & 98.2\% & 0.7\%  \\ \hline
          \% Wins   & 0.2\% & 99.3\% \\ \hline
        \end{tabular}

        \smallskip {\small (b) Without COI reduction}
      \end{center}
    \end{minipage}

    \smallskip
    \caption{Statistics summarizing the results on a universe of 450
      problem instances with a timeout of 3600s, with and without COI
      reduction.}
    \label{tab:stats}
  \end{center}
\end{table*}

\begin{table*}[t]
  \begin{center}
    \begin{minipage}{0.425\textwidth}
      \begin{center}
        \begin{tabular}{|c|c|c|} \hline
          \rowcolor{lightgray}
          EL2SAT & EL2MCS & Ties \\ \hline\hline
          2      & 19     & 429  \\ \hline
        \end{tabular}

        \smallskip {\small (a) With COI reduction}
      \end{center}
    \end{minipage}
    \begin{minipage}{0.125\textwidth}
      \quad
    \end{minipage}
    \begin{minipage}{0.425\textwidth}
      \begin{center}
        \begin{tabular}{|c|c|c|c|} \hline
          \rowcolor{lightgray}
          EL2SAT & EL2MCS & Ties \\ \hline\hline
          3      & 71     & 376  \\ \hline
        \end{tabular}

        \smallskip {\small (b) Without COI reduction}
      \end{center}
    \end{minipage}

    \smallskip
    \caption{Number of times a tool computes more MUSes within a
      timeout of 3600s, with and without COI reduction.}
    \label{tab:wins}
  \end{center}
\end{table*}
%
%
\section{Conclusions \& Future Work} \label{sec:conc}

Axiom pinpointing serves to identify unintended subsumption relations
in DLs. For the $\fml{EL}$ family of DLs, there has been recent work
on axiom pinpointing, the most efficient of which is based on encoding
the problem to propositional Horn formulae.
The main contribution of this paper is to relate axiom pinpointing
with MUS extraction. As a result, this enables exploiting different
MUS extraction and enumeration algorithms for the problem of axiom
pinpointing.
Preliminary results, obtained using off-the-shelve tools, show
categorical performance gains over the current state of the art in
axiom pinpointing for the $\fml{EL}$ family of DLs.

The results are justify further analysis of the uses of MUSes and
MCSes in axiom pinpointing, for the $\fml{EL}$ family of DLs, but also
for other DLs.
%

\subsubsection*{\ackname}
We thank the authors of EL2SAT for authorizing the use of the most
recent, yet unpublished, version of their work.
This work is partially supported by SFI PI grant BEACON (09/IN.1/\-I2618),
by FCT grant POLARIS (PTDC/EIA-CCO/\-123051/\-2010), and by national funds
through FCT with reference UID/CEC/50021/2013.


\end{document}